\documentclass[preprint,showkeys,secnumarabic,showpacs,amsmath,amssymb]{revtex4}
\usepackage{graphicx}
\begin{document}

\title{The curvaton scenario in Brane cosmology: model parameters and their constraints }

\author{Hossein Farajollahi}
\email{hosseinf@guilan.ac.ir} \affiliation{Department of Physics,
University of Guilan, Rasht, Iran}

\author{Arvin Ravanpak}
\email{aravanpak@guilan.ac.ir} \affiliation{Department of Physics,
University of Guilan, Rasht, Iran}

\date{\today}

\begin{abstract}
We have studied the curvaton scenario in brane world cosmology
in an intermediate inflationary scenario. This approach has allowed us to
find some constraints on different parameters that appear in the
model.
\end{abstract}



\maketitle

\section{Introduction}

Inflation as the most promising framework for
understanding the physics of the very early universe, ties the
evolution of the universe to the properties of one or more scalar
inflaton fields, responsible for creating an accelerating expanding universe.
This then creates a flat and homogeneous universe which later evolves into the
present universe. In general, the identity of the scalar inflaton
fields are currently unknown and lies outside the
standard model for particle physics \cite{Tzirakis}. It is well known that
inflation is also the most requiring solution to many other
deep-rooted problems of the cosmological models such as horizon
and flatness problems. Another success of the inflationary universe
model is that it provides a causal interpretation of the origin of
the observed anisotropy of the cosmic microwave background radiation (CMBR)
, and also the distribution of large scale structures
\cite{Campo}--\cite{Hinshaw}.

Intermediate inflation is a class of cosmological models where the scale factor
of the universe behaves in a certain extend intermediate between a power law
and exponential expansion \cite{Rendall} where the expansion rate
is faster than power law and slower than exponential ones \cite{Sanyal}--\cite{Muslimov}.
 In this class, the slow-roll
approximation conditions are well satisfied with time and therefor similar to
power-law inflation, there is no natural end to inflation within the
model \cite{Barrow}.

The problem of inflation in the Randall-Sundrum model of a single brane in
an AdS bulk successfully incorporates the idea
that our universe lies in a three-dimensional brane within a
higher-dimensional bulk spacetime \cite{Randall}. All the brane-world
inflationary models in the high energy limit posses correction terms
in their Friedmann equations. Although these terms have important
consequences in the inflationary dynamics, as the energy density
 decreases, these corrections become
unimportant, and the inflaton field enters a kinetic energy
dominated regime and brings inflation to an end. An alternative reheating
mechanism might be required since the inflaton may
survive this process without decay \cite{Papantonopoulos}. While during the inflationary period the
universe becomes dominated by the inflation scalar potential, at
the end of inflation the universe represents a combination of kinetic
and potential energies related to the scalar field, which is
assumed to take place at very low temperature \cite{Campuzano}.

In the standard reheating mechanism while the temperature grows in many orders of magnitude, most of the matter and
radiation of the universe was created, via the decay of the inflaton
field and the Big-Bang universe is recovered. This reheating temperature is of particular interest. In this
era the radiation domination, in which there exist a number
of particles of different kinds, begins. In the standard mechanism of reheating, the stage of
 oscillation of the scalar field is very fundamental in which a minimum in the
inflaton potential is something crucial for the reheating mechanism.
However, since the scalar field potential in these models do not present a minimum, the usual
mechanism introduced to bring inflation to an end becomes ineffective \cite{Kofman}.
These models are known in the literature as non-oscillating or simply NO Models \cite{Felder}.
To overcome this problem, one of the mechanism of
reheating in these kind of models is the introduction of the curvaton
field \cite{del}--\cite{Lyth}.

The curvaton scenario was in the first place suggested as an alternative
mechanism to generate the primordial scalar perturbation responsible for the
 structure formation \cite{Papantonopoulos}. Its decay into conventional matter offers an efficient
mechanism of reheating, and its field has the property whose energy
density is not diluted during inflation therefore the curvaton may be
responsible for some or all the matter content of the universe at the
present time. Alternatively, the large scale structure of the universe may also be
explained by the curvaton field. However, in here, in the context of brane intermediate inflation, we would like to present
the curvaton field in a mechanism to bring inflation to
an end and explain the reheating mechanism, similar to the work adopted by the
authors in \cite{del} and \cite{Liddle}.  In
\cite{del}, the authors have used the curvaton mechanism for intermediate inflation
in standard cosmology and in \cite{Liddle} the authors have studied steep inflation and not intermediate inflation model
in brane cosmology.

\section{The Model}

We consider a five-dimensional brane cosmology in which the
modified Friedmann equations are given by \cite{Shiromizu,Kamenshchik},
\begin{equation}\label{fried}
H^2=\kappa\rho_\phi[1+\frac{\rho_\phi}{2\lambda}]+\frac{\Lambda_4}{3}+\frac{\xi}{a^4},
\end{equation}

\begin{equation}\label{p}
2\dot{H}+3H^2=-3\kappa(\rho_{\phi}+p_{\phi})[1+\frac{\rho_{\phi}}{\lambda}]+3\kappa\rho_{\phi}
[1+\frac{\rho_{\phi}}{2\lambda}]+\Lambda_4-\frac{\xi}{a^4}
\end{equation}
where $H = \dot{a}/a$, $\rho_\phi$, $p_{\phi}$ and $\Lambda_4$  represent
respectively the Hubble parameter, the energy and pressure of the matter field confined to the
brane and the four-dimensional cosmological constant. In here we
assume $\kappa = 8\pi G/3 = 8\pi/(3m_p^2)$. The last term in equation (\ref{fried}) or (\ref{p}) is called the radiation term and represents
the influence of the bulk gravitons on the brane, with $\xi$ as
an integration constant. It has been noted in \cite{last-term}
 that while during inflation this term can be neglected, but it might play an important role at the beginning of inflation. The brane tension $\lambda$ which relates
the four and five-dimensional Planck masses via the expression $m_p
= \sqrt{3M_5^6/(4\pi\lambda)}$, is constrained by nucleosynthesis to
satisfies the inequality $\lambda
> (1MeV)^4$. In the following, for the inflation epoch, we assume that the universe in in the high energy
regime, i.e. $\rho_\phi\gg\lambda$.
We also assume that the four-dimensional cosmological constant is
vanished and just when the inflation begins, the last term in (\ref{fried}) and (\ref{p}) will rapidly
become unimportant which, in turn, leaves us with the following effective equations
\begin{equation}\label{efffried}
H^2\simeq\beta\rho_\phi^2,
\end{equation}
\begin{equation}\label{ph}
2\dot{H}+3H^2=-3\beta(\rho_{\phi}+2p_{\phi})\rho_{\phi},
\end{equation}
where $\beta = \kappa/(2\lambda)$, with dimension of $m_p^{-6}$.

We also assume that the inflaton field is confined to the brane, thus
its field equation is in the standard form:
\begin{equation}\label{scalarfield}
\ddot{\phi}+3H\dot{\phi}+V'=0,
\end{equation}
where $V(\phi)$ is the effective
scalar potential. The dot means derivative with respect to the
cosmological time and prime means derivative with respect to scalar
field $\phi$. In addition, the conservation equation is,
\begin{equation}
\dot{\rho_\phi}+3H(\rho_\phi+p_\phi)=0,
\end{equation}
where $\rho_\phi=(\dot{\phi}^2/2)+V(\phi)$, and
$p_\phi=(\dot{\phi}^2/2)-V(\phi)$.  In here, for convenience we also take units in which
$c={\hbar}=1$.

For intermediate inflationary universe
models, the exact solution can be found by assuming that the scale factor $a(t)$ expands as,
\begin{equation}\label{inter}
a(t)=\exp(t^f),
\end{equation}
where $f$ is a constant parameter with range $0<f<1$.

From equations (\ref{efffried}), (\ref{scalarfield}) and (\ref{inter}) the expressions for the scalar potential
$V(\phi)$ and the scalar field $\phi(t)$ are respectively
\begin{equation}
V(\phi)={\frac{2(1-f)}{9\beta}}[6f{(\frac{3\beta^{1/2}}{4(1-f)})^f}\phi^{2(f-1)}-\frac{(1-f)}{\phi^2}],
\end{equation}
and
\begin{equation}\label{phi}
\phi(t)=[{\frac{4(1-f)}{3\beta^{1/2}}}t]^{1/2}.
\end{equation}
Then, the Hubble parameter as a function of the inflaton field
becomes
\begin{equation}\label{h}
H(\phi)=f(\frac{3\beta^{1/2}}{4(1-f)})^{f-1}\phi^{-2(1-f)}.
\end{equation}
The form of the scale factor expressed in equation (\ref{inter}) also arises when
we solve the field equations in the slow roll approximation, where a
simple power law scalar potential is considered as
\begin{equation}
V(\phi)=f\beta^{(f-2)/2}[\frac{3}{4(1-f)}]^{f-1}\phi^{-2(1-f)}.
\end{equation}
The
solutions for $\phi(t)$ and $H(\phi)$ obtained with this potential
in the slow roll approximation are similar to those obtained in
the exact solution, expressed by (\ref{phi}) and (\ref{h}).
Note also that for this kind of potential a minimum does not exit.

The dimensionless slow roll parameters $\varepsilon$ and $\eta$ which are defined
by $\varepsilon\simeq V'^2/(3\beta V^3)$ and $\eta\simeq V''/(3\beta
V^2)$, respectively, in our case reduce to
\begin{equation}
\varepsilon\simeq\frac{4(1-f)^2}{3f\beta^{f/2}[\frac{3}{4(1-f)}]^{f-1}}\phi^{-2f},
\end{equation}
and
\begin{equation}
\eta\simeq\frac{2(1-f)(3-2f)}{3f\beta^{f/2}[\frac{3}{4(1-f)}]^{f-1}}\phi^{-2f}.
\end{equation}
The ratio between $\varepsilon$ and $\eta$ is
$\varepsilon/\eta=2(1-f)/(3-2f)$ and thus for $0<f<1$ , $\eta$ is always larger
than $\varepsilon$. Note also that $\eta$ reaches
unity earlier than $\varepsilon$ does. Therefore, one can represent the end of inflation is governed by the condition $\eta=1$ in place
of $\varepsilon=1$. From this condition, for the inflaton field at
the end of inflation we obtain
\begin{equation}\label{phie}
\phi_e=[\frac{2(1-f)(3-2f)}{3f\beta^{f/2}[\frac{3}{4(1-f)}]^{f-1}}]^{1/2f},
\end{equation}
where, the subscript $"e"$ is used to denote the end of the
inflationary period. Also, the number of the e-folds corresponding to the cosmological
scales, i.e. the number of remaining inflationary e-folds at the
time when the cosmological scale exits the horizon defines as
\begin{equation}\label{Nstar}
N_\ast=\int_{t_\ast}^{t_e}{H(t')dt'}=A[\frac{3\beta^{1/2}}{4(1-f)}]^f(\phi_e^{2f}-\phi_\ast^{2f}).
\end{equation}

\section{The Curvaton Field During The Kinetic Epoch}

By neglecting the term $V'$ in the field equation (\ref{scalarfield}) in comparison to the friction term
$3H\dot{\phi}$, the model begins a new period which is
called the kinetic epoch or kination. Hereafter, we will use
the subscript (or superscript) $"k"$, to label different quantities
at the beginning of this era. During the kination era
we have $\dot{\phi}^2/2 > V(\phi)$ which could be seen as a stiff
fluid since $p_\phi=\rho_\phi$. Also, at the beginning of the kination we assume the low energy limit, i.e. $\rho_\phi\ll\lambda$. In this regime two friedmann equations become:
\begin{equation}\label{lfired}
H^2=\kappa\rho_{\phi},
\end{equation}
\begin{equation}\label{pl}
2\dot{H}+3H^2=-3\kappa p_{\phi}.
\end{equation}
In the kinetic epoch, the field equations (\ref{fried}) and (\ref{scalarfield}) become,
$H^2=\kappa\dot{\phi}^2/2$ and $\ddot{\phi}+3H\dot{\phi}=0$, where
the second equation gives,
\begin{equation}
\dot{\phi}={\dot{\phi}}_k(\frac{a_k}{a})^3.
\end{equation}
Then, the energy density and Hubble parameter respectively become
\begin{equation}\label{rhophi}
\rho_\phi(a)=\rho_\phi^k(\frac{a_k}{a})^6,
\end{equation}
and
\begin{equation}\label{hk}
H(a)=H_k(\frac{a_k}{a})^3,
\end{equation}
where $H_k$ and $\rho_\phi^k$ are the values of the Hubble parameter
and energy density associated to the inflaton field at the beginning
of the kinetic epoch.

The curvaton field obeys the Klein-Gordon equation and in here we assume
that the scalar potential associated to this field is given by
$U(\sigma)=m^2\sigma^2/2$, with $m$ to be the curvaton mass.

We now assume that $\rho_\phi$ to be the dominant component when
compared to the curvaton energy density, $\rho_\sigma$. In
addition, the curvaton field oscillates around the minimum of its
effective potential $U(\sigma)$. During the kination, the
inflaton remains dominated and the curvaton density
evolves as a non-relativistic matter, i.e. $\rho_\sigma\propto
a^{-3}$. Then, the curvaton
field decay into radiation and the standard Big-Bang cosmology is
recovered.

In the inflation era, it is assumed that the curvaton field
is effectively massless. In the same period the curvaton rolls down
its potential until its kinetic energy is weakened by the
exponential expansion that its kinetic energy has almost vanished,
and so it becomes frozen. The curvaton field then assumes roughly to
be a constant, $\sigma_\ast\approx\sigma_e$, where, the
subscript $"\ast"$ refers to the era when the cosmological scale
exits the horizon.

In here, we also assume that during the kination the
Hubble parameter decreases so that its value is equivalent with the
curvaton mass, i.e. $m\simeq H$, where at this stage, the curvaton
field becomes effectively massive. Then, from equation (\ref{hk}), we
obtain
\begin{equation}\label{ratiom}
\frac{m}{H_k}=(\frac{a_k}{a_m})^3,
\end{equation}
where the subscript $"m"$ represents quantities at the time when the
curvaton mass $m$ during kination is of the order of $H$.

To prevent a period of curvaton-driven inflation the
universe must still be dominated by the inflaton field, i.e.
$\rho_\phi^m\gg\rho_\sigma$($\sim U(\sigma_e)\simeq
U(\sigma_\ast))$. This inequality permits us to find a constraint on
the values of the curvaton field $\sigma_\ast$. Now when
$H\simeq m$, we get $\frac{m^2\sigma_\ast^2}{2\rho_\phi^m}\ll1$,
which means that the curvaton field $\sigma_\ast$ satisfies the
constraint,
\begin{equation}\label{sigmaastsquare}
\sigma_\ast^2\ll\frac{2}{\kappa}\cdot
\end{equation}
At the end of inflation, the ratio of the potential energies
becomes,
\begin{equation}
\frac{U_e}{V_e}=\frac{m^2\sigma_\ast^2\beta^{1/2}}{2H_e}<1,
\end{equation}
and, thus, the curvaton energy becomes subdominant at the end
of inflation. The curvaton mass then should comply the constraint,
\begin{equation}\label{mh}
m^2<H_e^2=f^{2/f}(\frac{3-2f}{2})^{2(f-1)/f}.
\end{equation}
The constraint (\ref{mh}) imposed by the fact that the curvaton field must be
effectively massless during the inflationary era and thus $m<H_e$. When the mass of the curvaton field becomes important,
i.e. $m\simeq H$, its energy density decays like a
non-relativistic matter in the form of
$\rho_\sigma=m^2\sigma_\ast^2a_m^3/(2a^3)$. This is because, as the
curvaton undergoes quasi-harmonic oscillations, the
potential and kinetic energy densities become comparable.

The decay of the curvaton field may happen in two different
situations. One, when the curvaton field first dominates the cosmic
expansion and then decays. Two, when the decay of the curvaton field
occurs before it dominates the cosmological expansion. In the following section we will investigate these situations in more details.

\section{Constraints on model parameters}

{\bf Case 1: curvaton decay after domination}

If the curvaton field dominates the cosmic expansion (i.e.
$\rho_\sigma>\rho_\phi$), then, at a distance, (say $a=a_{eq}$), the
inflaton and curvaton energy densities must become equal. Therefore, from (\ref{rhophi}) and (\ref{hk}) and
using $\rho_\sigma\propto a^{-3}$, we obtain,
\begin{equation}\label{ratiophi}
(\frac{\rho_\sigma}{\rho_\phi})|_{a=a_{eq}}=\frac{\kappa
m^2\sigma_\ast^2}{2}\frac{a_m^3}{a_k^6}\frac{a_{eq}^3}{H_k^2}=1.
\end{equation}
in addition, from equations (\ref{hk}), (\ref{ratiom}) and (\ref{ratiophi}), we find a relation for the Hubble
parameter, $H_{eq}$, in terms of the curvaton parameters,
\begin{equation}\label{hableeq}
H_{eq}=H_k(\frac{a_k}{a_{eq}})^3=\frac{\kappa m\sigma_\ast^2}{2}.
\end{equation}
Since the decay parameter $\Gamma_\sigma$ is constrained by
nucleosynthesis, it is required that the curvaton field decays
before nucleosynthesis, i.e., $H_{nucl}\sim10^{-40}<\Gamma_\sigma$ (in units of Planck mass
$m_p$). By the requirement that $\Gamma_\sigma<H_{eq}$, we acquire a constraint on the decay
parameter $\Gamma_\sigma$, as
\begin{equation}\label{10-40}
10^{-40}<\Gamma_\sigma<\frac{\kappa m\sigma_\ast^2}{2}.
\end{equation}
Following the argument given by the authors in \cite{del}, we constrain the parameters appearing in
our model by revising the scalar perturbations related to the
curvaton field $\sigma$. At the time when the
decay of the curvaton field occurs, the Bardeen parameter, $P_\zeta$,
whose observed value is about $2.4\times 10^{-9}$\cite{Komatsu}, becomes \cite{Lyth},
\begin{equation}\label{bardeen0}
P_\zeta\simeq\frac{1}{9\pi^2}\frac{H_\ast^2}{\sigma_\ast^2}\cdot
\end{equation}
Since the spectrum of fluctuations is automatically gaussian for
$\sigma_\ast^2\gg H_\ast^2/(4\pi^2)$, and  also is independent of
$\Gamma_\sigma$, the analysis of space
parameter in our model is simplified. With the help of equations (\ref{mh}) and (\ref{10-40}) we obtain
\begin{equation}\label{gammaafter}
\Gamma_\sigma<\frac{\kappa\sigma_\ast^2}{2}f^{1/f}(\frac{3-2f}{2})^{(f-1)/f},
\end{equation}
that imposes an upper limit on $\Gamma_\sigma$.

Meanwhile, from Big Bang Nucleosynthesis $(BBN)$ temperature, $T_{BBN}$, we constrain our model parameter $f$.
The reheating occurs before $BBN$, with $T_{BBN}\sim10^{-22}$ (in unit of $m_p$), and thus we have
$T_{reh}>T_{BBN}$. By knowing that
$T_{reh}\sim\Gamma_\sigma^{1/2}>T_{BBN}$ we obtain the constraint,
\begin{equation}\label{hast}
H_\ast^2=f^2[\frac{3-2f}{2f}-N_\ast]^{2(f-1)/f}>(540\pi/8)^{2/3}
(P_\zeta T_{BBN}^2)^{2/3}\sim10^{-34},
\end{equation}
where, for the later, we have used the scalar spectral index $n_s$, closed to one $(m\approx H_*/10)$ \cite{Sanchez}.

In FIG.1 (left panel), the number of e-folds, $N_\ast$,
with respect to $f$ is shown by fitting the constraint (\ref{hast}) in its lower
limit. Alternatively, when the upper limit, $H_\ast\leqslant10^{-5}$\cite{Dimopoulos}, is used, the number of e-folds against the parameter $f$ is shown in the FIG.1 (right panel).

{\bf Case 2: curvaton decay before domination}

We assume that the curvaton decays before its energy
density becomes greater than the inflaton one. Moreover, the
mass of the curvaton is comparable with the Hubble
expansion rate, such that we can use $\rho_\sigma\propto
a^{-3}$. Following \cite{del}, we also assume that the curvaton field decays at a time when
$\Gamma_\sigma= H(a_d)=H_d$ and therefore from equation (\ref{hk}) we get,
\begin{equation}\label{gammasig}
\Gamma_\sigma=H_d=H_k(\frac{a_k}{a_d})^3,
\end{equation}
where "d" stands for quantities at the time when the curvaton
decays.

If we assume that the curvaton field decays after its mass
becomes important, (so that $\Gamma_\sigma<m$) and also before the
curvaton field dominates the expansion of the universe (i.e.,
$\Gamma_\sigma>H_{eq}$), we then obtain a new constraint as,
\begin{equation}\label{kappa}
\frac{\kappa\sigma_\ast^2}{2}<\frac{\Gamma_\sigma}{m}<1.
\end{equation}
Again, following \cite{del}, for this scenario, we find that the Bardeen parameter becomes,
\begin{equation}\label{bardeen}
P_\zeta\simeq\frac{r_d^2}{16\pi^2}\frac{H_\ast^2}{\sigma_\ast^2},
\end{equation}
where $r_d=\frac{\rho_\sigma}{\rho_\phi}|_{a=a_d}$. By using
$\rho_\sigma=m^2\sigma_\ast^2a_m^3/(2a^3)$ and equations (\ref{rhophi}), (\ref{ratiom}) and (\ref{gammasig})
we obtain,
\begin{equation}\label{r-d}
r_d=\frac{\kappa m\sigma_\ast^2}{2\Gamma_\sigma}.
\end{equation}
Now, From equations (\ref{mh}) and (\ref{kappa}), we get the inequality,
\begin{equation}\label{gammabefore}
\Gamma_\sigma<f^{1/f}(\frac{3-2f}{2})^{(f-1)/f},
\end{equation}
which gives an upper limit on $\Gamma_\sigma$. Finally, since the reheating temperature satisfies
$T_{reh}>T_{BBN}$, and also $\Gamma_\sigma>T_{BBN}^2$, we derive a new constraint as
\begin{equation}\label{hast2}
H_\ast^2=f^2[\frac{3-2f}{2f}-N_\ast]^{2(f-1)/f}>(120\pi)^{2/3}
(P_\zeta T_{BBN}^2)^{2/3}\sim10^{-34},
\end{equation}
where, as before, we have used the scalar spectral index $n_s$
closed to one. Note that apart from the coefficient, the expression is similar to the one
obtained for the decay of curvaton field after domination, as
expressed by equation (\ref{hast}). Therefore, the graph of the number of e-folds, $N_*$, against $f$ is similar to the one obtained in case of the carvaton decay after domination.

\begin{figure}[h]
\centering
\includegraphics[scale=0.35]{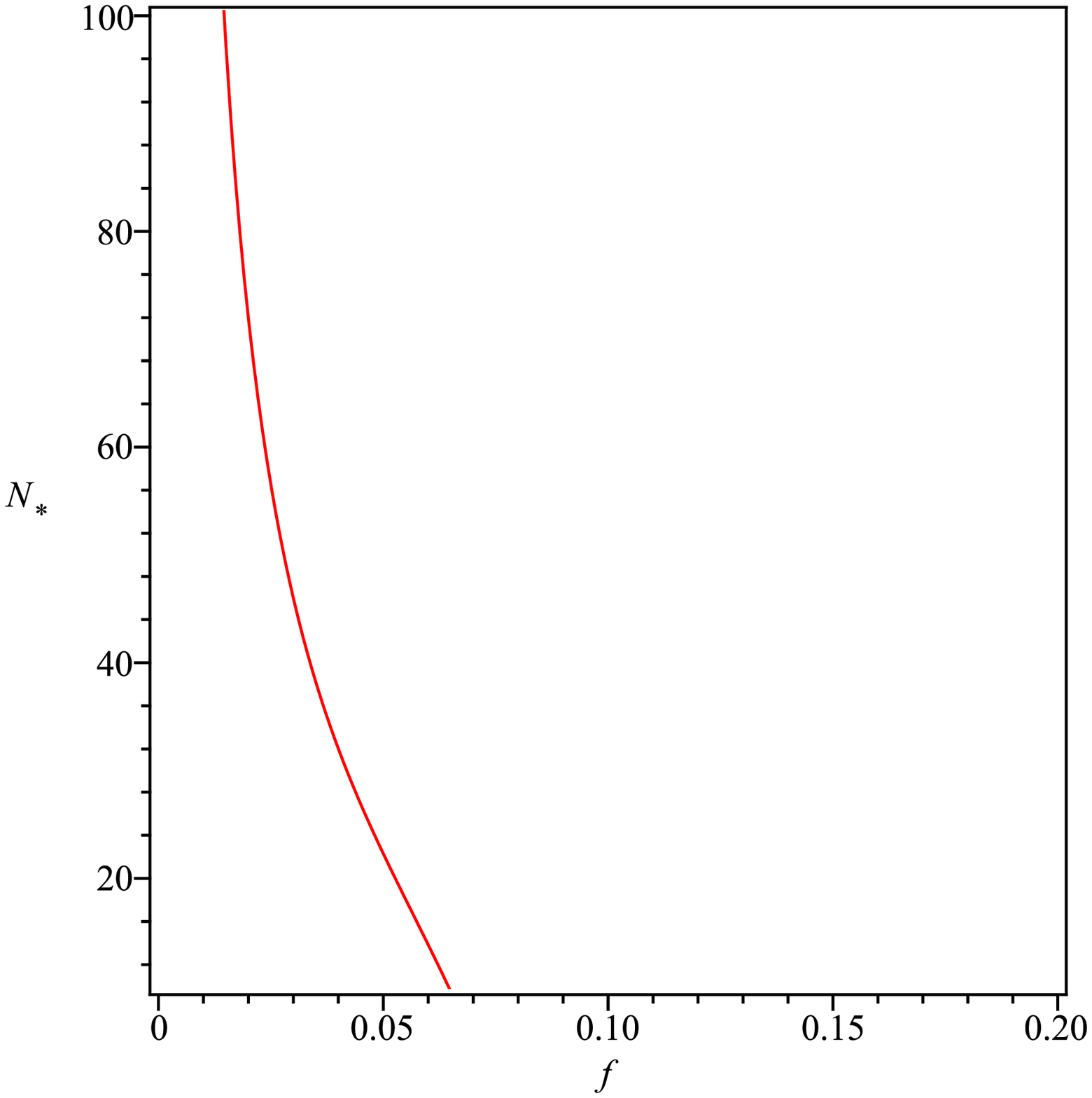}\includegraphics[scale=0.35]{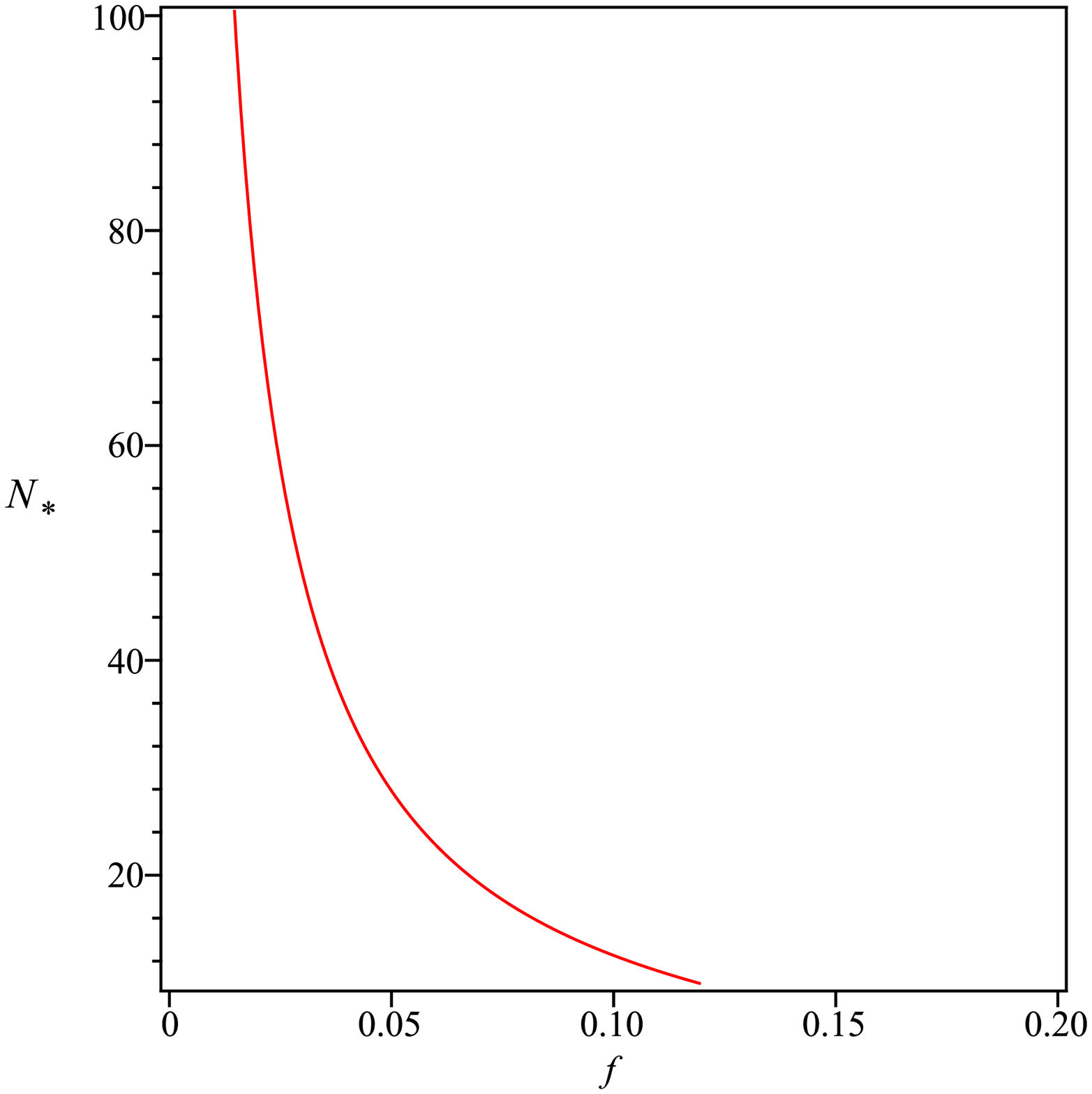}
\caption{The number of the e-folds with respect to $f$ in both curvaton decay after and before domination, fitted from the lower limit of the $T_{BBN}$ (left panel) and upper limit of the
$T_{BBN}$ (right panel).}
\end{figure}

\section{Constraints on Curvaton Mass}

Similar to the case of constraint on $\Gamma_\sigma$
parameter, one may constrain on the value of the curvaton mass, using tensor perturbation methods. In these methods, the corresponding
gravitational wave amplitude can be written as \cite{Kamenshchik},
\begin{equation}\label{hgw}
h_{GW}\simeq C_1H_\ast,
\end{equation}
where $C_1$ is an arbitrary constant. Therefore, one may take $H\ll10^{-5}$, which means that
inflation may occur at an energy scale smaller than the grand
unification. This, in turn, is an advantage of the curvaton approach in
comparison with the single inflaton field one.

From the modified Friedmann equation we have
$H_\ast^2=\beta V_\ast^2$, and thus by using equation (\ref{hast2}) we obtain,
\begin{equation}\label{grav-wave}
h_{GW}^2\simeq C_1^2f^2[\frac{3-2f}{2f}-N_\ast]^{2(f-1)/f}.
\end{equation}
From equations (\ref{mh}) and (\ref{grav-wave}), we yield the inequality
\begin{equation}
m^2<\frac{h_{GW}^2}{C_1^2}[1-\frac{2fN_\ast}{3-2f}]^{2(1-f)/f},
\end{equation}
which imposes an upper limit to the curvaton mass.

From \cite{Langlois}, the constraint on the density fraction of the gravitational
wave is given by
\begin{equation}\label{I}
I\equiv
h^2\int_{k_{BBN}}^{k_\ast}\Omega_{GW}(k)dlnk\simeq2h^2\epsilon\Omega_\gamma(k_0)h^2_{GW}
(\frac{H_\ast}{\widetilde{H}})^{2/3}\leq2\times 10^{-6},
\end{equation}
where $k_{BBN}$ and $\Omega_{GW}(k)$ are respectively the physical
momentum corresponding to the horizon at $BBN$ and the density fraction of the gravitational
wave with physical momentum $k$. The density fraction
of the radiation at present on horizon scales
is $\Omega_\gamma(k_0)=2.6\times10^{-5}h^{-2}$. Also,
$\epsilon\sim10^{-2}$ and $h=0.73$ is the Hubble constant in which
$H_0$ is in units of $100 km/sec/Mpc$. The parameter $\widetilde{H}$
is either $\widetilde{H}=H_{eq}$ or $\widetilde{H}=H_d$, for the curvaton decays
after or before domination.

{\bf Case 1: curvaton decay after domination}

In this case, using  equations (\ref{hableeq}), (\ref{bardeen0}) and (\ref{hgw}),
the constraint on the density fraction of the gravitational wave,
expressed by equation (\ref{I}), becomes
\begin{equation}\label{m-sigma}
\frac{m}{\sigma_\ast^2}\gtrsim
2.5\times10^{-13}{P_\zeta}^2\sim10^{-30}.
\end{equation}
A second constraint is obtained by incorporating (\ref{10-40}) and (\ref{m-sigma}):
\begin{equation}\label{msquare0}
m^2>5.96\times10^{-54}P_\zeta^2\simeq3.43\times10^{-71}
\end{equation}
If the decay rate is of gravitational strength, then, $\Gamma_\sigma\sim m^3$ \cite{Rodriguez}--\cite{Lazarides}
and equation (\ref{10-40}) imposes a third constraint as
\begin{equation}\label{msquare1}
\frac{m^2}{\sigma_\ast^2}<\frac{4\pi}{3}\simeq4.19.
\end{equation}
From the previous sections, we also have the constraints (\ref{sigmaastsquare}) and (\ref{10-40}). One may also find another constraint by using
equations (\ref{mh}) and (\ref{bardeen0}) on the curvaton mass for the curvaton decay after domination as
\begin{equation}\label{ec}
\frac{m^2}{\sigma_\ast^2}<9\pi^2P_\zeta\sim 2.13\times10^{-7}.
\end{equation}
Altogether, there are six constraints (\ref{sigmaastsquare}), (\ref{10-40}), (\ref{m-sigma})--(\ref{ec}) on either $m$ or $\sigma_\ast$, for the
curvaton decay after domiantion. In FIG. 2(left panel), the allowed region for curvaton mass against curvaton field is shown in shaded color.

{\bf Case 2: curvaton decay before domination}

In case 2, the constraint on the density fraction of the gravitational
wave, expressed by equation (\ref{I}), with regards to $\Gamma_\sigma^{1/2}>T_{BBN}$, becomes,
\begin{equation}\label{m-sigmastar}
m\sigma_\ast>3.2\times10^{-4}P_\zeta^{1/2}T_{BBN}^{3/2}\sim10^{-41},
\end{equation}
by using equations (\ref{gammasig}), (\ref{bardeen}) and (\ref{r-d}). A second constraint by incorporating (\ref{kappa}) and (\ref{m-sigmastar}) is given by,
\begin{equation}\label{msquare}
m^2>4.3\times10^{-7}P_\zeta T_{BBN}^3\simeq1.04\times10^{-81}.
\end{equation}
By using equation (\ref{kappa}), a pair of new constraints can be obtained as,
\begin{equation}\label{msquare2}
m^2<1,
\end{equation}
and
\begin{equation}\label{msquare3}
\frac{m^2}{\sigma_\ast^2}>4.19.
\end{equation}
Furthermore, from equation (\ref{m-sigmastar}) one can get another constraint as,
\begin{equation}\label{sigmam5/4}
\frac{\sigma_\ast}{m^{5/4}}>3.2\times10^{-4}P_\zeta^{1/2}\simeq1.57\times10^{-8}.
\end{equation}
The constraint (\ref{sigmaastsquare}) can also be imposed on the curvaton  field in the case of curvaton decay before domination.
In addition, one may find another constraint by using (\ref{mh}), (\ref{kappa}), (\ref{bardeen}) and (\ref{r-d}) as,
\begin{equation}\label{ec2}
m^2\sigma_\ast^2<9P_\zeta\sim 2.16\times10^{-8}.
\end{equation}
There are seven constraints (\ref{sigmaastsquare}), (\ref{m-sigmastar})--(\ref{ec2}) on either $m$ or $\sigma_\ast$ if the curvaton decay before domiantion. In Fig. 2(right panel), The shaded region shows the allowed region for $m$ which is bounded by the above constraints.

\begin{figure}[h]
\centering
\includegraphics[scale=0.35]{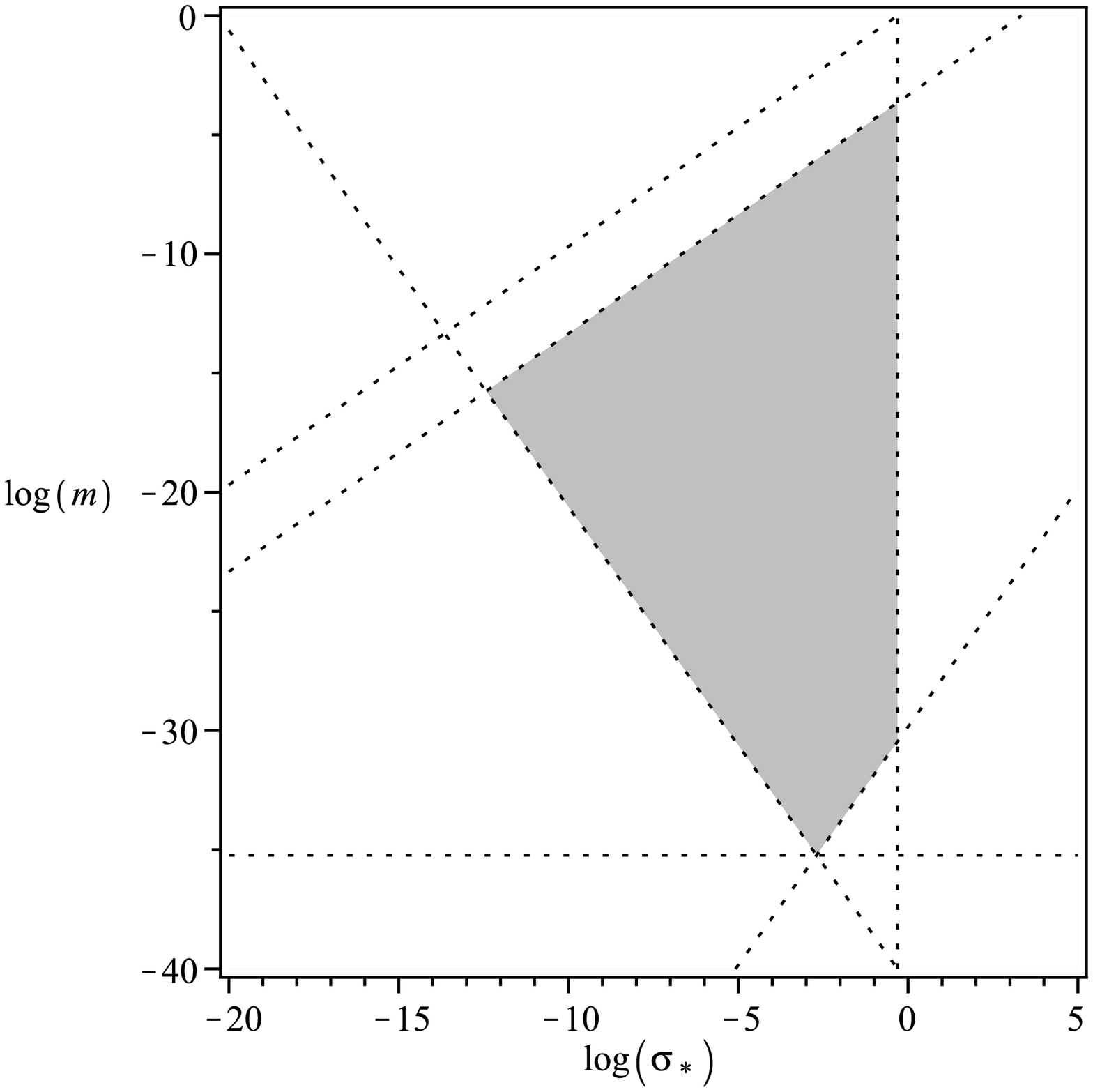}\includegraphics[scale=0.35]{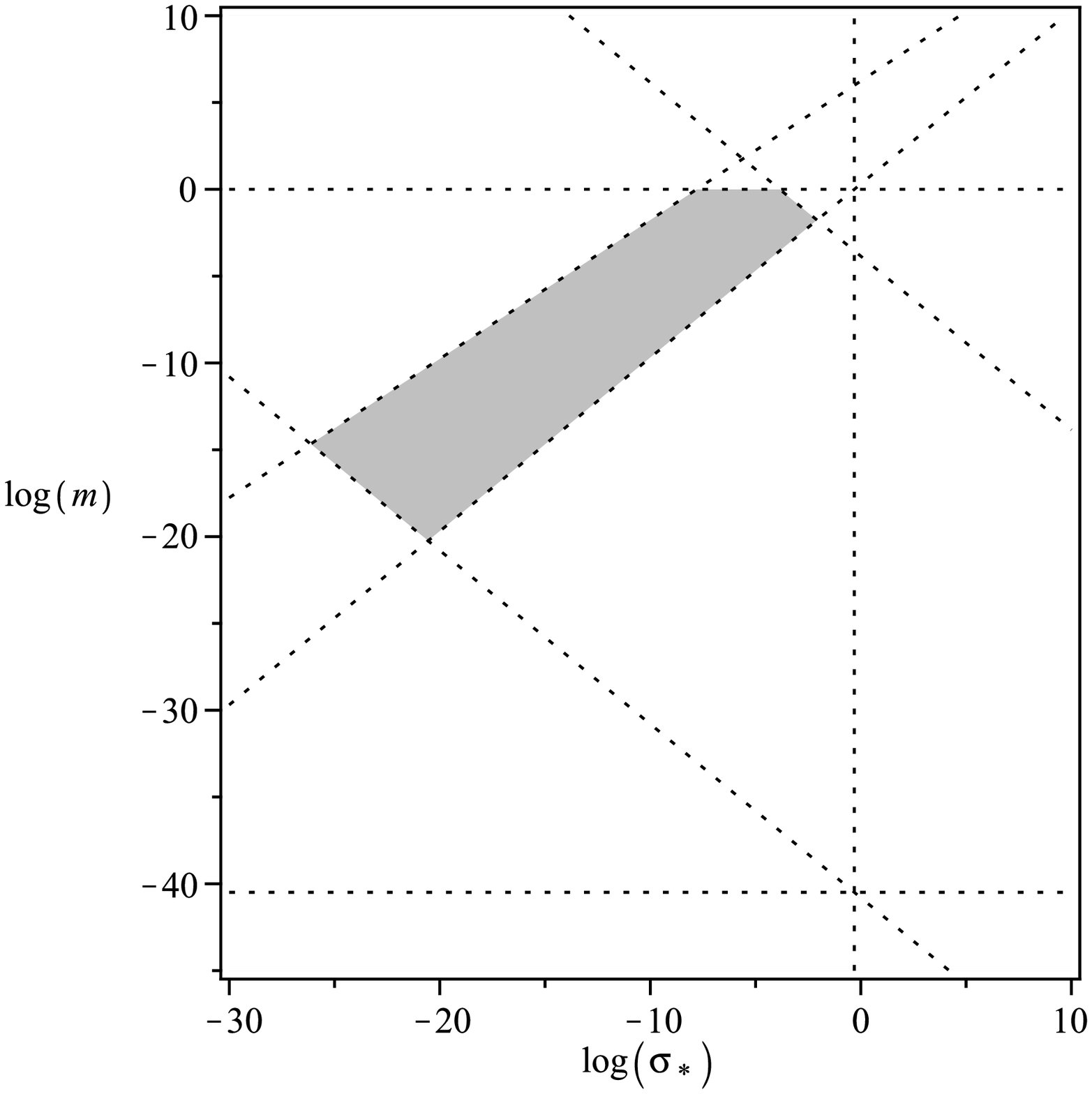}
\caption{Allowed region of parameter space of the curvaton-brane intermediate inflation
model for the case of curvaton domination after decay (left panel) and curvaton domination before decay (right panel). The allowed regions are
shaded.}
\end{figure}

{\bf constraint on the reheating temperature}

In a very limited case, another constraint can be imposed on the decay rate of the curvaton field as \cite{Nojiri},
\begin{equation}
\Gamma_\sigma=g^2m,
\end{equation}
where $g$ is the coupling of the curvaton to its decay products. Then, the
allowed range for the coupling constant is given by,
\begin{equation}\label{max}
max(\frac{T_{BBN}}{m^{1/2}},m)\lesssim g \lesssim
min(1,\frac{m\sigma_\ast^3}{T_{BBN}^2}),
\end{equation}
where the inequality $m\lesssim g$ is due to gravitational decay. For
the first case and using $T_{reh}>T_{BBN}$ this
constraint gives an upper limit given by $g<m\sigma_\ast^3/T_{BBN}^2$,
and when the curvaton decays after domination, a lower limit is
given by $T_{BBN}m^{-1/2}<g$.

By using equation (\ref{max}) and by considering $H_\ast\simeq10^{-17}$, $\sigma_\ast\sim10^{-12}$ \cite{Sanchez}
and $m\sim10^{-18}$ (from $n_s\simeq1$), we obtain that
$10^{-13}\lesssim g \lesssim10^{-10}$. As a result, since
$T_{reh}\sim gm^{1/2}$, the allowed range for the reheating
temperature becomes $10^{-22}\lesssim T_{reh}\lesssim10^{-19}$(in
units of $m_p$).

Alternatively, by choosing $\sigma_\ast=1$ \cite{Sanchez}, from expression (\ref{max}) the range for $g$ becomes
$10^{-13}\lesssim g\lesssim1$. Therefore, with $T_{reh}\sim gm^{1/2}$, the
allowed range for the reheating temperature becomes
$10^{-22}\lesssim T_{reh}\lesssim10^{-9}$(in units of $m_p$). The
constraints on the density fraction of gravitational waves suggest
$g\sim1$\cite{Sanchez}. Thus, we obtain that the reheating temperature
becomes of the order of $T_{reh}\sim10^{-9}$ (in units of $m_p$)
which challenges gravitino constraint \cite{Dine}.

\section{Summary}

We have analysed the scenario of intermediate
inflation in the brane-world cosmology in the presence of curavaton field. The curvaton
 field, responsible for universe reheating, imposes constraints on the model parameters.

For the intermediate inflationary universe with the scale factor given by equation (\ref{inter}), there is one space parameter $f$. Following \cite{del}, two possible scenarios have been taken to study universe reheating via curvaton mediation, where the curvaton dominates the universe after or before it decays. As a result, we have obtained an upper limit constrain for $\Gamma_\sigma$ expressed by equation (\ref{gammaafter}) in the first scenario. We have also found a lower limit constraint for the value of $\Gamma_\sigma$ which is represented
by equation (\ref{gammabefore}) for the second scenario. In both scenarios, we have acquired constraints for the parameters expresses by the equations (\ref{hast}) and (\ref{hast2}). The plot for different values of e-folding with respect to $f$ fitted from both lower and upper limits of $T_{BBN}$ is also given in FIG. (1).

For the curvaton field and its mass we have also obtained constraints in both scenarios. In the first scenario, there are six constraints which bound variation of curvaton mass $m$ with respect to the curvaton field $\sigma_\ast$ to the region shown in FIG. 2(left panel). Similarly, one can find the seven constraints limited these parameters and plotted in FIG. 2(right panel). Finally, we probe the reheating scenario in our model and come across the possible constraints on the reheating temperature by talking into account the result obtained from the previous sections.


\begin{thebibliography}{99}
\bibitem{Tzirakis}K. Tzirakis and W. H. Kinney, JCAP. 01, 028 (2009).

\bibitem{Campo}S. del Campo and R. Herrera, Phys. Lett. B670, 266-270 (2009).

\bibitem{3}A. Guth, Phys. Rev. D23, 347 (1981).

\bibitem{4}A. Albrecht and P. J. Steinhardt, Phys. Rev. Lett. 48, 1220
(1982).

\bibitem{5}J. Dunkley, E. Komatsu, M. R. Nolta, D. N. Spergel, D. Larson,
G. Hinshaw, L. Page, C. L. Bennett, B. Gold, N. Jarosik, J. L.
Weiland, M. Halpern, R. S. Hill, A. Kogut, M. Limon, S. S. Meyer, G.
S. Tucker, E. Wollack, E. L. Wright, Astrophys. J. Suppl. 180,
306-329 (2009).

\bibitem{Hinshaw}G. Hinshaw, J. L. Weiland, R. S. Hill, N. Odegard, D. Larson,
C. L. Bennett, J. Dunkley, B. Gold, M. R. Greason, N. Jarosik, E.
Komatsu, M. R. Nolta, L. Page, D. N. Spergel, E. Wollack, M.
Halpern, A. Kogut, M. Limon, S. S. Meyer, G. S. Tucker, E. L.
Wright, Astrophys. J. Suppl. 180, 225-245 (2009).

\bibitem{Rendall}A. D. Rendall, Class. Quant. Grav. 22, 1655-1666 (2005).

\bibitem{Sanyal}A. K. Sanyal, Adv. High Energy Phys. 2009, 630414 (2009) .

\bibitem{9}J. D. Barrow, Phys. Lett. B235, 40 (1990).

\bibitem{Muslimov}A. G. Muslimov, Class. Quantum. Grav. 7, 231 (1990).

\bibitem{Barrow}J. D. Barrow and A. R. Liddle, Phys. Rev. D47, 5219-5223 (1993).

\bibitem{Randall}L. Randall and R. Sundrum, Phys. Rev. Lett. 83, 3370-3373 (1999).

\bibitem{Papantonopoulos}E. Papantonopoulos and V. Zamarias, JCAP. 0611, 005 (2006).

\bibitem{Campuzano}C. Campuzano, S. del Campo and R. Herrera, Phys. Rev. D72, 083515 (2005).

\bibitem{Kofman}L. Kofman and A. Linde, JHEP. 0207, 004 (2002).

\bibitem{Felder}G. Felder, L. Kofman and A. Linde, Phys. Rev. D60, 103505
(1999).

\bibitem{del}S. del Campo and R. Herrera, Phys. Rev. D76, 103503 (2007).

\bibitem{18}S. Mollerach, Phys. Rev. D42, 313 (1990).

\bibitem{Lyth}D. H. Lyth and D. Wands, Phys. Lett. B524, 5 (2002).

\bibitem{Liddle}A. R. Liddle and L. A. Urena-Lopez, Phys. Rev. D68, 043517 (2003).

\bibitem{Shiromizu}T. Shiromizu, K. Maeda and M. Sasaki,
Phys. Rev. D62, 024012 (2000).

\bibitem{Kamenshchik}A. Kamenshchik, U. Moschella and V. Pasquier,
Phys. Lett. B511, 265 (2001).

\bibitem{last-term} S. del Campo, R. Herrera, V. Cardenas, PLB 672, 89 (2009)

\bibitem{Komatsu}E. Komatsu et al.[WMAP Collaboration], Astrophys. J. Suppl. 180, 330 (2009).

\bibitem{Sanchez}J. C. Bueno Sanchez, K. Dimopoulos, JCAP. 0711, 007 (2007).

\bibitem{Dimopoulos}K. Dimopoulos and D. H. Lyth, Phys. Rev. D69, 123509 (2004).

\bibitem{Langlois}D. Langlois, R. Maartens and D. Wands, Phys. Lett. B489, 259 (2000).

\bibitem{Rodriguez}K. Dimopoulos, D. H. Lyth and Y. Rodriguez, JHEP. 0502, 055 (2005).

\bibitem{28}K. Dimopoulos, D. H. Lyth, A. Notari and A. Riotto, JHEP. 0307, 053 (2003).

\bibitem{29}K. Dimopoulos, Phys. Lett. B634, 331 (2006).

\bibitem{Lazarides}K. Dimopoulos, G. Lazarides, D. Lyth and R. Ruiz de Austri, Phys. Rev. D68, 123515 (2003).

\bibitem{Herrera}C. Campuzano, S. del Campo, R. Herrera, E. Rojas and J. Saavedra,
Phys. Rev. D80, 123531 (2009).

\bibitem{Nojiri}S. Nojiri, S. D. Odintsov and M. Sasaki, Phys. Rev. D71, 123509 (2004).

\bibitem{Dine}M. Dine, Cambridge University Press,
First published in print format (2006).

\end{thebibliography}
\end{document}